# CCTFv1: Computational Modeling of CyberTeam Formation Strategies

Tristan J. Calay[1], Basheer Qolomany[2], Aos
Mulahuwaish[1], Liaquat Hossain[2], and Jacques Bou Abdo[3]

[1] Saginaw Valley State University, University Center, USA
[2] University of Nebraska at Kearney, Kearney, USA
[3] University of Cincinnati, Cincinnati, USA

**Abstract.** Rooted in collaborative efforts, cybersecurity spans the scope of cyber competitions and warfare. Despite extensive research into team strategy in sports and project management, empirical study in cybersecurity is minimal. This gap motivates this paper, which presents the Collaborative Cyber Team Formation (CCTF) Simulation Framework. Using Agent-Based Modeling, we delve into the dynamics of team creation and output. We focus on exposing the impact of structural dynamics on performance while controlling other variables carefully. Our findings highlight the importance of strategic team formations, an aspect often overlooked in corporate cybersecurity and cyber competition teams.

**Keywords:** Cybersecurity · Cyber Competition · Collaborative Teams

## 1 Introduction

Teams, more than just collectives engaged in mutual tasks, present complex dimensions studied across various fields, including organizational sociology, anthropology, organizational behavior, industrial psychology, sociometry, and social network analysis [8, 17, 31]. Echoing Aristotle's philosophy that the whole is greater than the sum of its parts, comprehending teams is key to understanding the evolution and future trajectory of industrial revolutions [29]. Their potential manifests in enhanced performance, improved quality through error identification and recovery [28], increased productivity via labor distribution, and fostering innovation [25]. As such, optimizing team performance becomes a significant factor for the success of organizations, enterprises, large-scale endeavors, and competitive teams [9, 11].

While teamwork can bolster performance, it may also introduce errors and dysfunction [28]. Factors such as power dynamics, leadership, culture, organizational processes, communication, cohesion, and team composition influence team performance [18]. This was demonstrated in FIFA's 2002 World Cup when the 1998 winning French team, despite retaining over 80% of its members [2, 3], and led by the same captain, Zinedine Zidane, failed to win any match. This downfall is attributed to a change in team formation from a defensive 4-3-3 in 1998 to a more balanced 4-2-3-1 in 2002, catching opponents off guard [1, 4]. Such strategies of role configuration are ubiquitous in professional team formation and have been extensively studied in competitive sports and project management [10, 21]. However, such exploration is scant in the cybersecurity domain.

This study aims to fill this gap and hypothesizes that team formation strategies significantly affect performance and victory probability. Utilizing computational modeling, we simulated a Blue team/Red team cyber competition, where teams were given two unique roles. The performance was assessed based on role configuration. To eliminate skill biases, we developed roles using Agent-Based Modeling, focusing on the impact of team formation strategies and network properties on performance.

This work serves as an initial stride in comprehending the performance of cyber teams, laying the groundwork for the development of the Collaborative Cyber Team Formation (CCTF). CCTF is a three-pillared framework that we are building to dissect the performance of cyber teams. The first pillar of CCTF is rooted in theoretical research that explores collaborative systems, particularly through the lenses of game theory, social learning, and complex systems. The second pillar hinges on the empirical analysis of cyber competitions coupled with subject matter expertise. The third pillar, to which this work contributes, draws on computational modeling.

## 2    Implementation Methodology

This work aims to assess the influence of team formation strategies on performance in cyber competitions, considering several influencing factors:

- **Team Size:** Larger teams tend to yield greater collective performance [23], indicating a direct proportionality between team size and performance.
- **Skills:** Teams with higher skills, encompassing talent [19] and training [27], typically outperform those with lesser skills, suggesting a direct correlation between members' skills and team performance.
- **Collaboration:** More collaborative teams generally perform better [20], implying a direct relationship between collaboration and performance. However, talent and collaboration may sometimes be inversely proportional [30].
- **Leadership:** Teams exhibiting superior leadership and trust usually perform better than those facing leadership crises [13].
- **Social Influence:** Performance can be swayed by social pressures within the team, like interpersonal dynamics [22], or external pressures, such as crowd influence [16]. This encompasses factors like motivation, stress, and discrimination.

To ascertain the impact of team formation distinct from these factors, the study design neutralizes or, when impractical, fixes these influences as detailed in Section 2.1. Moreover, performance is contingent upon specific goal attainment, thus, it is evaluated based on key predefined objectives, as elaborated in Section 2.1.

## 2.1   Study Design

In our study, a cyber competition is modeled as a defender team safeguarding a network against an attacker team. Following the models established in [14, 15], the network is constituted of interconnected nodes, with only the peripheral nodes (internet-facing) being accessible to attackers initially. On successfully overtaking a peripheral node, attackers can target linked nodes. Figure 1(a) depicts such a network, where peripheral nodes in the yellow region are exposed to attack attempts (black arrows). Non-peripheral nodes in the green region are unreachable directly but become accessible (blue arrows) once a peripheral node is compromised (red arrow).

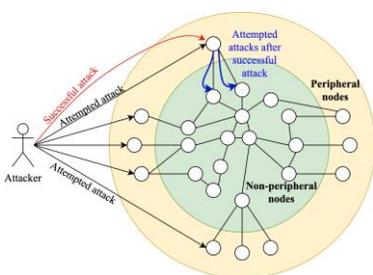

| Attacker Roles | Hacking Steps as defined by EC-Council | Incident Handling Steps as defined by EC-Council | Defender Roles |
|---|---|---|---|
| Role 2: Exploiter | Step 5: Clearing Track | | |
| | Step 4: Maintaining Access | Step 4: Act | Role 2: Interceptor |
| | Step 3: Gaining Access | Step 3: Decide | |
| Role 1: Scout | Step 2: Scanning | Step 2: Orient | Role 1: Detector |
| | Step 1: Reconnaissance | Step 1: Observe | |

(a) Modeling the network and attackers

(b) Attacker and defender roles matched to EC-Council's hacking and incident handling steps

**Fig. 1:** Modeling study agents

In our model, each attacker assumes one of two distinct roles: a "scout" or an "exploiter". The scout scans accessible nodes to detect vulnerabilities, with the probability of successful detection—given a node is vulnerable—set by the user and defined as:

$$P\,(detect \mid node\ is\ vulnerable) = P_{scout}$$

Once a scout identifies a vulnerability, exploiters are notified to attempt to exploit this vulnerability and control the compromised node. The probability of successful exploitation—given a node is vulnerable—is set by the user as:

$$P\,(exploit \mid node\ is\ vulnerable) = P_{exploiter}$$

Systematically, the scout performs the initial two hacking steps as defined by EC-Council [5], as illustrated in Figure 1(b). In parallel, the exploiter is systematically characterized as the role executing steps 3 through 5. The attacker team consists of $s$ scouts and $N\_s$ exploiters, where $N$ denotes the total team size. To mitigate the influence of "team size," as outlined in section 2, this study maintains constant and equivalent team sizes for both attacker and defender teams.

Agent-Based Modeling (ABM) is employed to curtail the impact of "skills", "collaboration", and "social influence" (detailed in Section 2.2). Identical agents mitigate "skills", while broadcasting communication and anonymous interac- tions limit "collaboration" and "social influence", respectively. An ad-hoc team arrangement, with a user-determined size ($N$), helps manage the "leadership" factor. Success in a cyber-attack hinges on its objectives; this study outlines three unique objectives and corresponding metrics (Table 1).

**Table 1:** Measured Objectives

| Index | Objective | Metric |
|---|---|---|
| 1 | Overtake as many nodes as possible | Portion of the network overtaken by the attackers. This metric is a decimal number ranging between 0 and 1. |
| 2 | Overtake the whole network | Whether the whole network was overtaken by the at- tackers. This metric is a Boolean. |
| 3 | Overtake the central nodes repre- senting core services such as the database | Whether the central nodes were overtaken by the at- tackers. This metric is a Boolean. |

In our model, each defender can assume one of two distinct roles: a "detector" or an "interceptor". The detector scans the network for compromised (exploited or overtaken) or vulnerable nodes. The probability of successfully identifying an infected node—given that it is indeed infected—is set by the user and defined as:

$$P(detect \mid node\ is\ infected) = P_{detector-exploited}$$

while the probability of successfully identifying a vulnerable node—given that it is indeed vulnerable—is set by the user and defined as:

$$P(detect \mid node\ is\ infected) = P_{detector-vulnerable}$$

Upon detection of an infected node, interceptors are alerted to defend the net- work. The interceptor can either flag the infected node as untrusted—thus iso- lating it [12, 24, 26]—or recover the node based on user-defined action. The in- terceptor requires time $\Delta_{interceptor}$, set by the user, to carry out this action. The detector is systematically aligned with the first three steps of incident han- dling, as defined by EC-Council [6], as illustrated in Figure 1(b). Similarly, the exploiter is systematically defined as the role that performs step 4. The team of defenders is composed of $d$ detectors and $N\ d$ interceptors. The same measures implemented to limit the impact of the factors defined in section 2 on attacker team's performance are implemented to limit the impact on defender team's performance.

**Hypothesis** As noted in section 2.1, team performance, gauged by the three metrics, is influenced by the team formation strategies of both attackers and defenders. In this section, we posit that team formation strategies notably impact performance and the likelihood of victory. Given that team formation invariably precedes performance and potential confounding factors have been controlled, correlation analysis remains necessary for deciding whether to accept or reject the proposed hypothesis.

## 2.2 Study Parameters

Our methodology utilizes an Agent-Based Modeling approach featuring five agent types: Router, Scout, Exploiter, Detector, and Interceptor. These are illustrated in our NetLogo simulation, as shown in Figure 2. While the network size and structure can be user-specified, for the rest of this paper, we are considering a scale-free network with 30 routers. Each input parameter can accommodate a broad spectrum of values; however, for demonstration, we have used incremental values iteratively as enumerated in Table 2.

It is notable that every tick triggers packet generation and routing table broadcasts, with the simulation offering real-time status updates to emulate a realistic network model. The workflow, outlined in Figure 3, involves offline routers becoming active again once the simulation surpasses their shutdown delay. Meanwhile, scouts and detectors can identify vulnerable routers, exploiters have the capacity to compromise routers, and interceptors can both shut down and rehabilitate routers.

**Table 2:** Input Parameters

| Input Parameter | Description | Simulation Values |
|---|---|---|
| $N$ | Total number of team members ($N$) for each side. | 10 |
| $S$ | Number of attacker scouts ($S$). $1 \leq S < N$. Exploiters: $N - S$. | 1 - 9 |
| $d$ | Number of defender detectors ($d$). $1 \leq d < N$. Interceptors: $N - d$. | 1 - 9 |
| $Vul\_rate$ | Chance of a router becoming vulnerable when vulnerabilities are generated. | 2% |
| $P_{scout}$ | Probability of attacker scout discovering a vulnerable router. | 100% |
| $P_{exploiter}$ | Probability of exploiter to exploit an exploitable router. | 2% |
| $P_{detector-vulnerable}$ | Probability of defender detector discovering a vulnerable router. | 25%, 50%, 75%, 100% |
| $P_{detector-exploited}$ | Probability of defender detector discovering an exploited router. | 25%, 50%, 75%, 100% |
| $\Delta_{interceptor}$ | Time needed to restore a router. | 10 ticks |

# 3 Results and Analysis

Our study accommodates 1296 unique input combinations and corresponding outputs. Given the inherently stochastic nature of the simulation, we performed five trials for each setup to capture possible output variations. Each result is evaluated against the three objectives, as described in section 2.1, employing three established metrics. As a result, we derived a dataset comprising 6480 distinct combinations and corresponding outcomes. This dataset, along with the source code [7] for reproducibility and validation, is publicly accessible. The results are represented graphically in section 3.1. We subsequently posit and scrutinize a hypothesis using this assembled dataset in section 3.2.

## 3.1 Sample Results

In this section, we showcase the derived dataset via graphical representation. All diagrams, as depicted in figures 4 and 5, employ the number of Exploiters

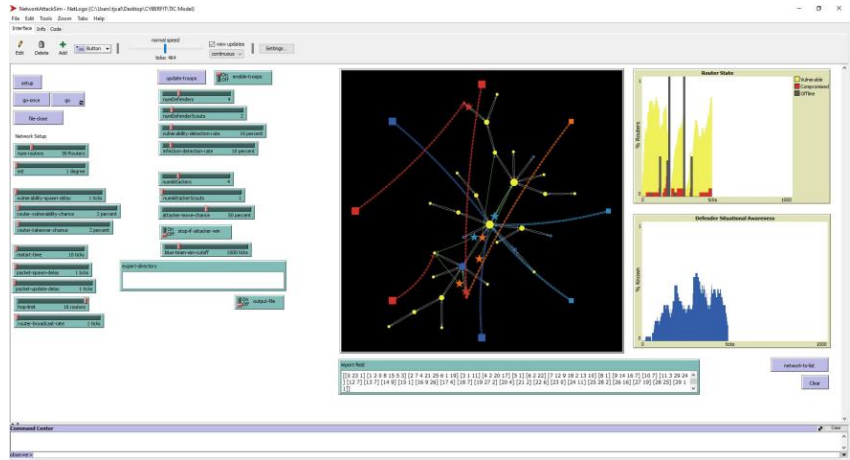

**Fig. 2:** NetLogo simulation snapshot

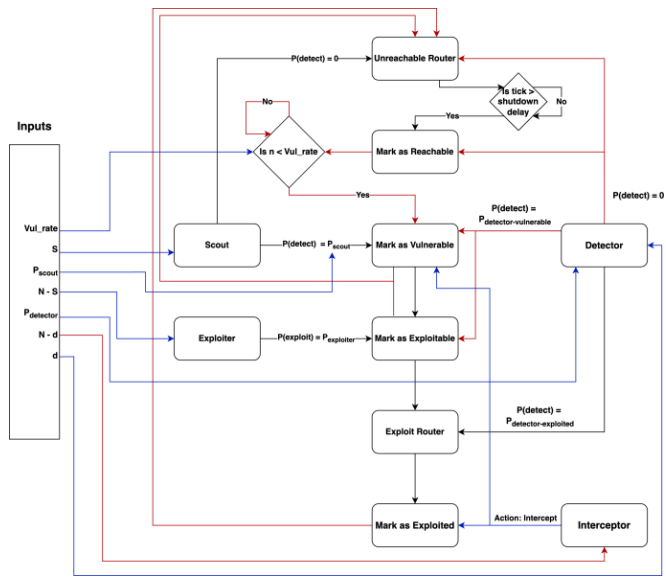

**Fig. 3:** NetLogo simulator workflow diagram

($N - s$) as the y-axis and the number of Interceptors ($N - d$) as the x-axis. The z-axis corresponds to the utilized metric for all configurations and tests.

A router is deemed compromised from the instant it is exploited until recovery initiation by an interceptor. Similarly, a router is classified as offline from the commencement of its recovery (or that of its root router) until successful recuperation. Tracking both compromised and offline routers is paramount, as the former represents the infiltrated network and the success of the attackers, while the latter signifies the network's containment and the defenders' success in thwarting the attack. However, both are critical for "Metric 1" (defined in section 2.1). Figure 4(a) illustrates the mean percentage of the network compro- mised throughout the test, while Figure 4(b) displays the maximum percentage. Likewise, Figures 4(c) and 4(d) represent the mean and maximum percentages of the offline network over the test duration, respectively. These four figures all utilize the same axes: number of interceptors ($N\ d$), number of exploiters ($N$ s), and network percentage. These visuals collectively indicate that, per metric 1, the defenders' optimum performance is achieved with a formation of 8 detectors and 2 interceptors. However, this team structure doesn't consistently yield superior performance in other configurations.

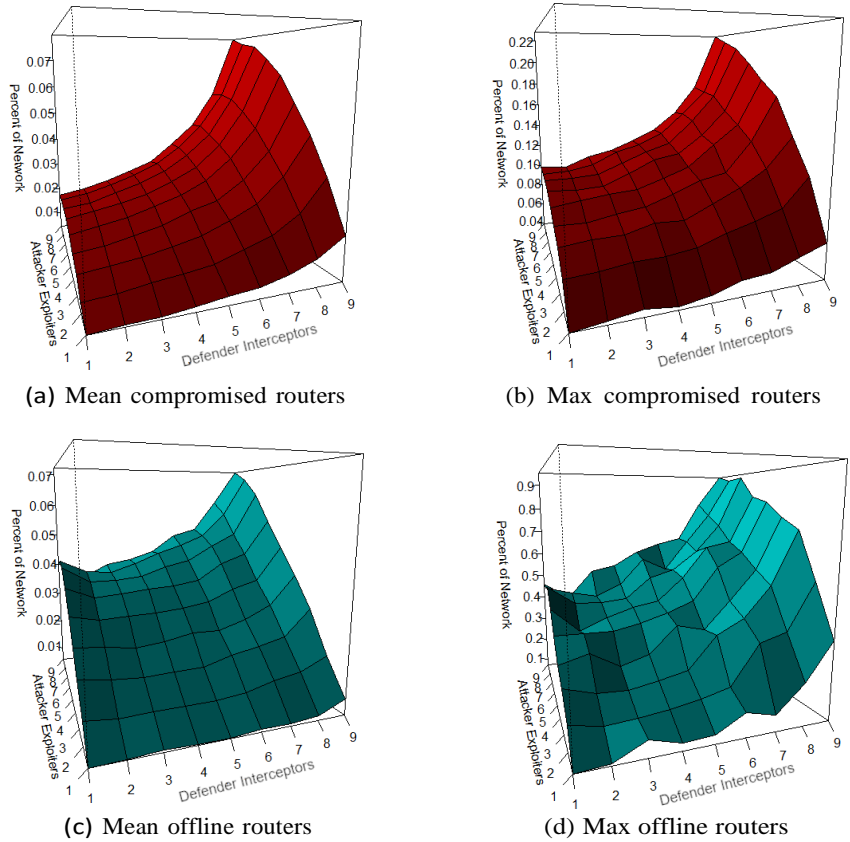

(a) Mean compromised routers

(b) Max compromised routers

(c) Mean offline routers

(d) Max offline routers

**Fig. 4:** Team performance measured following "Metric 1"

In accordance with metrics 2 and 3, depicted in figures 5(a) and 5(b), re- spectively, the defenders' formation of 8 detectors and 2 interceptors does not consistently exhibit top performance. The ideal formation for defenders is con-

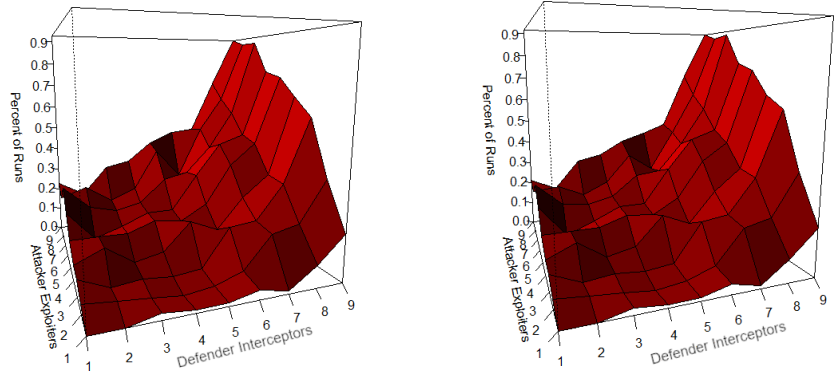

(a) Team performance following "Metric 2", percentage of the runs that resulted in the attackers compromising more than two-thirds of the network

(b) Team performance following "Metric 3", percentage of the runs that resulted in the attackers compromising the central router

**Fig. 5:** Team performance measured following "Metric 2" and "Metric 3"

tingent on the attackers' formation, an outcome anticipated from a game theory standpoint. It's also important to acknowledge that these results are based on the sample configuration and are furnished for illustrative purposes. A more comprehensive analysis will follow.

### 3.2   Hypothesis Testing

In section 2.1, we hypothesized that team formation strategies impact performance and the likelihood of victory. To test this hypothesis, we compute the correlation coefficient (R values) to gauge the strength of the relationship between two variables, thereby allowing us to understand and evaluate their statistical association independent of any extraneous variables. Table 3 presents the correlation between the performance metrics (defined in section 2.1, extracted from the generated dataset, and depicted in figures 4 and 5) and the Attacker/De- fender strategies. In this study, our hypothesis is validated, as there exists a significant negative correlation between the strategies of the attackers/defenders and the designated metrics. As the quantity of Attacker exploiters or Defender interceptors escalates, we observe a corresponding decline in metrics, and vice versa.

**Table 3:** Performance Metrics

| R Values | Mean Compro-mised | Max Compro-mised | Mean Offline | Max Of-fline | 2/3 Net-work (Metric 2) | Center Wins (Metric 3) |
|---|---|---|---|---|---|---|
| Attacker Strategy | -0.47 | -0.54 | -0.74 | -0.43 | -0.28 | -0.28 |
| Defender Strategy | -0.67 | -0.54 | -0.27 | -0.39 | -0.36 | -0.36 |

## 4    Conclusion

In this paper, we proposed that the formation strategy plays a crucial role in the performance of cyber teams. To investigate this hypothesis, we adopted a computational approach, and the results yielded significant findings that support the acceptance of the hypothesis. Moving forward, our next endeavor involves validating and potentially calibrating this computational model using empirical data collected from a competition. To ensure the integrity of the study, the competition will be organized as a controlled experiment, effectively minimizing the influence of other factors that could introduce biases. This work serves as a foundational component in the development of the Collaborative Cyber Team Formation (CCTF) framework, which aims to integrate theoretical, empirical, and computational approaches to comprehensively understand the performance dynamics of cyber teams.

## References

1. France 1998-2006, https://footballsgreatest.weebly.com/france-1998-2006.html
2. French national team 1998, https://www.fifa.com/tournaments/mens/worldcup/1998france/teams/43946
3. French national team 2002, https://www.fifa.com/tournaments/mens/worldcup/2002korea-japan/teams/43946
4. How france really won the world cup, https://www.aspeninstitute.org/blog-posts/how-france-really-won-the-world-cup/
5. What is ethical hacking?, https://www.eccouncil.org/cybersecurity/what-is-ethical-hacking/
6. What is incident response?, https://www.eccouncil.org/cybersecurity/what-is-incident-response/
7. Collaborative Cyber Team Formation (CCTF) Simulation Framework. githubrepository (2023), https://github.com/Starwhip/CCTF-Framework
8. Bar, M., Kempf, A., Ruenzi, S.: Is a team different from the sum of its parts? evidence from mutual fund managers. Review of Finance **15**(2), 359–396 (2011)
9. Bavelas, A.: A mathematical model for group structures. Human organization **7**(3), 16–30 (1948)
10. Budak, G., Kara, İ., İç, Y.T., Kasımbeyli, R.: New mathematical models for team formation of sports clubs before the match. Central European Journal of Opera- tions Research **27**, 93–109 (2019)
11. Cadima, R., Ojeda Rodríguez, J., Monguet Fierro, J.M.: Social networks and per- formance in distributed learning communities. Educational technology and society**15**(4), 296–304 (2012)
12. Callaway, D.S., Newman, M.E., Strogatz, S.H., Watts, D.J.: Network robustness and fragility: Percolation on random graphs. Physical review letters **85**(25), 5468 (2000)
13. Dirks, K.T.: Trust in leadership and team performance: Evidence from ncaa bas- ketball. Journal of applied psychology **85**(6), 1004 (2000)

14. Dobson, G.B., Carley, K.M.: Cyber-fit: an agent-based modelling approach to simulating cyber warfare. In: Social, Cultural, and Behavioral Modeling: 10th International Conference, SBP-BRiMS 2017, Washington, DC, USA, July 5-8, 2017, Proceedings 10. pp. 139–148. Springer (2017)
15. Dobson, G.B., Carley, K.M.: Cyber-fit agent-based simulation framework version 4. Center for the Computational Analysis of Social and Organizational Systems (2021)
16. Friesen, A.P., Wolf, S.A., van Kleef, G.A.: The social influence of emotions within sports teams. Feelings in sport pp. 49–57 (2020)
17. Greco, L.M., Porck, J.P., Walter, S.L., Scrimpshire, A.J., Zabinski, A.M.: A meta-analytic review of identification at work: Relative contribution of team, organizational, and professional identification. Journal of Applied Psychology 107(5), 795 (2022)
18. Grove, J.R., Fish, M., Eklund, R.C.: Changes in athletic identity following team selection: Self-protection versus self-enhancement. Journal of applied sport psychology 16(1), 75–81 (2004)
19. Gula, B., Vaci, N., Alexandrowicz, R.W., Bilalić, M.: Never too much—the benefit of talent to team performance in the national basketball association: Comment on swaab, schaerer, anicich, ronay, and galinsky (2014). Psychological Science 32(2), 301–304 (2021)
20. Harris, C.M., McMahan, G.C., Wright, P.M.: Talent and time together: The impact of human capital and overlapping tenure on unit performance. Personnel Review 41(4), 408–427 (2012)
21. Jahanbakhsh, F., Fu, W.T., Karahalios, K., Marinov, D., Bailey, B.: You want me to work with who? stakeholder perceptions of automated team formation in project-based courses. In: Proceedings of the 2017 CHI conference on human factors in computing systems. pp. 3201–3212 (2017)
22. Johnson, A.R., Van de Schoot, R., Delmar, F., Crano, W.D.: Social influence interpretation of interpersonal processes and team performance over time using bayesian model selection. Journal of Management 41(2), 574–606 (2015)
23. Mao, A., Mason, W., Suri, S., Watts, D.J.: An experimental study of team size and performance on a complex task. PloS one 11(4), e0153048 (2016)
24. Moore, C., Newman, M.E.: Epidemics and percolation in small-world networks. Physical Review E 61(5), 5678 (2000)
25. Osburg, T., Schmidpeter, R.: Social innovation. Solutions for a sustainable future p. 18 (2013)
26. Sahimi, M.: Applications of percolation theory, vol. 213. Springer Nature (2023)
27. Salas, E., DiazGranados, D., Klein, C., Burke, C.S., Stagl, K.C., Goodwin, G.F., Halpin, S.M.: Does team training improve team performance? a meta-analysis. Human factors 50(6), 903–933 (2008)
28. Sasou, K., Reason, J.: Team errors: definition and taxonomy. Reliability Engineering & System Safety 65(1), 1–9 (1999)
29. Senge, P.M., Carstedt, G., Porter, P.L.: Next industrial revolution. MIT Sloan management review 42(2), 24–38 (2001)
30. Swaab, R.I., Schaerer, M., Anicich, E.M., Ronay, R., Galinsky, A.D.: The too-much-talent effect: Team interdependence determines when more talent is too much or not enough. Psychological Science 25(8), 1581–1591 (2014)
31. Xu, H., Bu, Y., Liu, M., Zhang, C., Sun, M., Zhang, Y., Meyer, E., Salas, E., Ding, Y.: Team power dynamics and team impact: New perspectives on scientific collaboration using career age as a proxy for team power. Journal of the Association for Information Science and Technology 73(10), 1489–1505 (2022)